\documentclass[aps,preprint]{revtex4}
\usepackage{graphicx}
\usepackage{latexsym}
\usepackage{amssymb}
\usepackage{amsmath}
\usepackage{amsfonts}
\usepackage{calc}

\begin{document}

\title{Liquid Nanodroplets Spreading on Chemically Patterned Surfaces}
\author{Gary S. Grest}
\author{David R. Heine$^{\dagger}$}
\author{Edmund B. Webb III$^{\ast}$}
\affiliation{Sandia National Laboratories,  Albuquerque NM 87185 \\
$^{\dagger}$Present address - Corning Incorporated, Painted Post, New York 14870 \\
$^{\ast}$Author to whom correspondence should be addressed; email:  ebwebb@sandia.gov}

\date{\today}

\begin{abstract}
Controlling the spatial distribution of liquid droplets on surfaces via
surface energy patterning
can be used to control material delivery
to specified regions via selective liquid/solid wetting.
While studies of the equilibrium
shape of liquid droplets on
heterogenous substrates exist, much less is known about
the corresponding wetting kinetics.
We make significant progress towards elucidating details of this
topic by studying, via large-scale atomistic simulations,
liquid nanodroplets spreading on chemically patterned surfaces.
A model is presented for lines of polymer liquid
(droplets) on substrates
consisting of alternating strips of wetting (equilibrium
contact angle $\theta_0 \simeq 0^{\circ}$)
and non-wetting ($\theta_0 \simeq 90^{\circ}$) material.
Droplet spreading
is compared for different wavelength $\lambda$ of the pattern and
strength of surface interaction on the wetting strips.  For
small $\lambda$, droplets
partially spread on both the wetting and non-wetting regions of
the substrate to attain a finite contact angle
less than $90^{\circ}$.  In this case,
the extent of spreading depends
on the interaction strength in the wetting
regions.  A transition is observed such that, for large $\lambda$,
the droplet spreads only on the wetting region of the substrate
by pulling material from non-wetting regions.  In most cases, a
precursor film spreads on the wetting portion of the substrate
at a rate strongly dependent upon interaction
strength.
\end{abstract}

\maketitle

\section{Introduction\label{sec:intro}}
The spreading of a liquid droplet on a surface proceeds
as the droplet balances the interfacial tensions between the solid,
liquid, and vapor phases.  On a homogenous substrate, completely
wetting droplets (i.e. where
the equilibrium contact angle $\theta_0 = 0^{\circ}$)
often form a monolayer film or precursor foot on the surface that spreads
ahead of the main droplet.  This has been observed in experiments
\cite{HFC:Nat:89,HCL:PRL:89,HCL:PRL:90,XSB:PRL:04}
as well as simulation \cite{NAK:PRL:92,OCK:PRE:96,BKV:PRL:96,HGW:PRE:03,HGW:PRE:04}.
This foot grows diffusively with an effective diffusion constant
$D_{\rm eff}$ that depends on the square root of
the initial radius $R_o$ of the droplet \cite{HGW:PRL:05}.
While much has been revealed about the kinetics of droplets
spreading on homogenous surfaces,
spreading on chemically patterned surfaces is much less understood.
In particular the influence of the pattern feature size on
the spreading rate of both the precursor foot and main
droplet is unknown.

With the advent of modern lithography techniques, it is now possible to chemically
pattern a substrate with lyophobic and lyophilic surface domains
on the nano to
micron  scale \cite{XW:ARMS:98}.
These chemically
structured surfaces can be useful in  controlling the wetting of liquid droplets.
Depending on the size of
the pattern and the contact angle in the two domains,
wetting transitions can occur in which
the liquid droplet can change its shape or morphology 
\cite{Lip:COCIS:01,DT:ARFM:05}.
The equilibrium shapes of liquid droplets on a variety of
chemically patterned surfaces 
has been studied both experimentally and theoretically
\cite{GHLL:SCI:99,DTMW:JAP:00,BL:JAP:02,LDBY:LAN:03,DY:LAN:05,YMB:JCP:05,DPR:JOP:05}.
Introduction of a lyophobic grid can be used to confine liquid  droplets
and has been shown to  improve the quality of ink jet printing
\cite{DLBY:CM:05}.   
In most of these studies, the liquid droplet is
confined by the pattern of lyophilic and lyophobic domains and does not spread.

An alternate situation occurs if one places
a long liquid line transverse to a chemically striped surface of
alternating lyophilic
and lyophobic  domains.
This is the situation we study here for the case of
liquid polymer droplets of varying chain length.
For small $\lambda$, the main droplet partially wets
both regions of the surface equivalently and
does not spread even in
the wetting region.
For sufficiently large $\lambda$
the main droplet spreads on the wetting region of the substrate
by pulling material from non-wetting regions.  In most cases, a
precursor film spreads on the wetting portion of the substrate
at a rate strongly dependent upon interaction
strength even when the main droplet does not spread.

\section{Computational Procedure}\label{sec:comp}
Here we present extensive molecular dynamics (MD) simulations of the spreading of
short chain polymer droplets spreading on a chemically patterned surface.  The polymers
are modeled by a coarse grained bead spring model in which each polymer contains $N$
beads per chain with $N=10$ and $100$. As
the entanglement length $N_e\simeq 72$ for this model \cite{PKG:EPL:00},
the chains are unentangled.
The polymer is represented by spherical beads
of mass $m$ attached by springs, which interact with a truncated
Lennard-Jones (LJ) potential,

\begin{equation}
U_{LJ}(r)=\left\{ \begin{array}{rl}
4\varepsilon\left[\left(\frac{\sigma}{r}\right)^{12}-\left(\frac{\sigma}{r}\right)^{6}\right] & r\leq r_{c}\\
0 & r>r_{c}\end{array}\right.
\label{eq:ljcut}
\end{equation}
where $\varepsilon$ and $\sigma$ are the LJ units of energy and
length and the cutoff $r_{c}=2.5\:\sigma$. The monomer-monomer interaction
$\varepsilon$ is used as the reference and all monomers have the
same diameter $\sigma$. For bonded monomers, we apply an additional
potential where each bond is described by the FENE potential \cite{KG:JCP:90}
with $k=30\:\varepsilon/\sigma^{2}$ and $R_{0}=1.5\:\sigma$. The
substrate is modeled as a flat surface since it was found previously
\cite{HGW:PRE:03} that with the proper choice of thermostat,
simulations using a flat surface exhibit the same behavior as a realistic
atomic substrate with greater computational efficiency.

The interactions
between the surface and the monomers in the droplet at a distance
$z$ from the surface are modeled using an integrated LJ potential,

\begin{equation}
U_{LJ}^{wall}(z)=\left\{ \begin{array}{rl}
\frac{2\pi\varepsilon_{w}}{3}\left[\frac{2}{15}\left(\frac{\sigma}{z}\right)^{9}-\left(\frac{\sigma}{z}\right)^{3}\right] & z\leq z_{c}\\
0 & z>z_{c}\end{array}\right.
\label{eq:ljwall}
\end{equation}
where $\varepsilon_{w}$ is the monomer-surface interaction strength
and $z_{c}=2.2\sigma.$
Extending the range of this surface interaction to infinity shifts
the wetting transition to a lower
energy; however, the qualitative spreading
behavior is identical to the $z_{c}=2.2\sigma$
case \cite{HGW:PRL:05}.
For $z_c=2.2\sigma$, the critical wetting
strength (i.e. $\theta_0 \rightarrow 0^{\circ}$)
is $\varepsilon_w^c\simeq 1.75\varepsilon$
for $N=10$ and $\varepsilon_w^c\simeq 2.25\varepsilon$ for $N=100$ \cite{HGW:PRE:04}.
The system is periodic in
the $y$ direction with length $L_{y}$ and open in the other two directions.
The chemical pattern consists of
infinite strips of wetting and non-wetting regions of equal width $\lambda/2$
as shown in Figs.~1-2 and Fig.~4.
In the present
simulations the monomer/wall interaction $\varepsilon_w$
depends on the position of each monomer in the $y$ direction.
For the non-wetting region, $\varepsilon_w$ was
chosen so that $\theta_0 \simeq 90^o$, which for $N=10$ is
$\varepsilon_w=0.9\varepsilon$ and for $N=100$ is
$\varepsilon_w=1.0\varepsilon$ \cite{HGW:PRE:04}.   In the wetting region,
two cases were studied for $N=10$: $\varepsilon_w=2.0\varepsilon$
and $\varepsilon_w=3.0\varepsilon$, corresponding to weak and
strong wetting, respectively.  For $N=100$, $\varepsilon_w=2.5\varepsilon$, which is
in the weak wetting regime.
All of the droplets presented here are modeled as hemicylinders as
described previously \cite{HGW:PRE:04} with
initial droplet radii $R_{0}\cong50\sigma$.
The effect of varying $R_0$ for homogeneous droplets has been studied previously
\cite {HGW:PRL:05}.

To provide a realistic representation
of the transfer of energy in the polymer droplet, a
Langevin thermostat is applied only near the surface,

\begin{equation} 
m_{i}\frac{d^{2}\mathbf{r_{i}}}{dt^{2}}=-\Delta U_{i}-m_{i}\gamma_{L}\frac{d\mathbf{r_{i}}}{dt}+\mathbf{W}_{i}(t),
\label{eq:lang}
\end{equation} 
where $m_{i}$ is the mass of monomer $i$, $\mathbf{r_{i}}$ is the position
of monomer $i$, $\gamma_{L}$ is the 
friction coupling term for the Langevin thermostat, $-\Delta U_{i}$ is
the force acting on monomer $i$ due to the potentials defined above,
and $\mathbf{W}_{i}(t)$ is a Gaussian white noise term.
We use a Langevin coupling term with a damping
rate that decreases exponentially away from the substrate \cite{BP:PRE:01},
$\gamma_{L}(z)=\gamma_{L}^{s}\exp\left(\sigma-z\right)$.
Here $\gamma_{L}^{s}=3.0\tau^{-1}$ and $z$
is the distance from the substrate. 
The effect of varying $\gamma_L^s$ 
for homogeneous surfaces has been discussed in ref.~\cite{HGW:PRE:03}.
By including only a coupling of monomers near the surface to the thermostat,
we avoid the unphysical
effect of screening the hydrodynamic interactions in the droplet.
We also damp monomers near the surface stronger than those in the
bulk, which is a more physically reasonable description of dissipation
mechanisms due to the surface interaction.

In the simulations $L_y$ varied from $60\sigma$ to
$200\sigma$, depending on the periodicity of the pattern $\lambda$.
For $\lambda\le 60\sigma$, $L_y=60\sigma$ while for $\lambda\ge 100\sigma$,
$L_y=\lambda$. Results for a homogenous wetting substrate, labeled
$\lambda=\infty$ are for $L_y=60\sigma$.
The total size of the droplets varied from 
$N\cong200\,000$ to $680\,000$ monomers depending on the value of $L_y$.
The vapor pressure is low so that spreading
does not occur via vaporization and condensation.
However for small $\lambda$,
particularly for $N=10$, a few chains separate from the precursor foot and randomly
diffuse
on to the non-wetting region of the substrate and sometimes evaporate from the substrate.
These chains are subsequently removed from the simulation once they move away from
the surface a distance  several times $R_0$.

The equations of motion are integrated using a velocity-Verlet algorithm
with a time step of $\Delta t=0.01\;\tau$
where $\tau=\sigma\left(\frac{m}{\varepsilon}\right)^{1/2}$. The
simulations are performed at a temperature $T=\varepsilon/k_{B}$
using the \textsc{lammps} code \cite{P:JCP:95}, modified to include the
chemically patterned surface.  The instantaneous contact
radius for the main droplet $r_{b}(t)$ and
the radius of the precursor foot $r_{f}(t)$
are determined every $400\tau$
according to the procedure described previously \cite{HGW:PRE:04}. 
Briefly, the system is divided into layers of thickness $1.5\sigma$
parallel to the interface.  Data
presented for foot kinetics are obtained from
the layer closest to the surface (layer $1$) and data for bulk kinetics are
obtained from layer $4$, which is far enough above the substrate
to provide an accurate description of bulk droplet behavior.

\section{Results}\label{sec:results}

Time sequences for $N=10$ and $N=100$ are shown in Fig.~1 for $\lambda=200\sigma$.
For the main droplet to spread (bulk spreading),
the gain in interface energy must exceed the cost
of increasing the surface area of the droplet.
This balance depends on $\lambda$, $N$, and $\varepsilon_w$.
For $N=10$, bulk spreading occurs on the lyophilic strip as
material is transfered from the non-wetting to the wetting region.
As such, the cross-section
of material in the non-wetting region shrinks while maintaining a hemi-cylinder
shape with contact angle near $90^o$.  At later time in the
simulation, material is completely removed from non-wetting
regions.  Given periodicity in the $y$ direction, this
demonstrates a case where an intial line of liquid (in $y$)
transforms into a series of parallel liquid strips (in $x$) with
separation distance in $y$ equal to $\lambda/2$.
For $N=100$, evidence of $N$ dependence is observed in that bulk
spreading does not occur on the wetting region.
A precursor film or foot is observed to
spread ahead of the main droplet on the lyophilic strip
in both systems.
Thus, even when the main droplet does not spread,
a precursor foot may spread on the wetting region of the substrate.
For all cases where a foot was observed to spread,
the contact radius follows $r_{f}^{2}(t)=2D_{\rm eff}t$
where $D_{\rm eff}$ is the effective diffusion coefficient.
For $N=10$ and $\lambda=200\sigma$ (Fig.~1),
the effective diffusion constant $D_{\rm eff}$ of the precursor foot is
essentially equal to that for the pure wetting case $\lambda=\infty$.
Spreading of the precursor foot for $N=100$
causes depletion of material on the
wetting region.  Furthermore, bulk material is unable
to move quickly enough from the non-wetting region to the wetting
region, resulting in the formation of a saddle
shape at the center of the droplet.
At late times,
the precursor film spreads by polymer moving perpendicularly
from the non-wetting
to the
wetting region, then continuing along the wetting region. In this case
$D_{\rm eff}$ is $40\%$ of that for $\lambda=\infty$.  Although we do
not access such
time scales in our simulations, we expect that material will completely
wick away from the non-wetting region as the precursor film continues
to spread.

\begin{figure}
\begin{center}
\includegraphics[%
  clip,
  width=15.0cm,
  keepaspectratio]{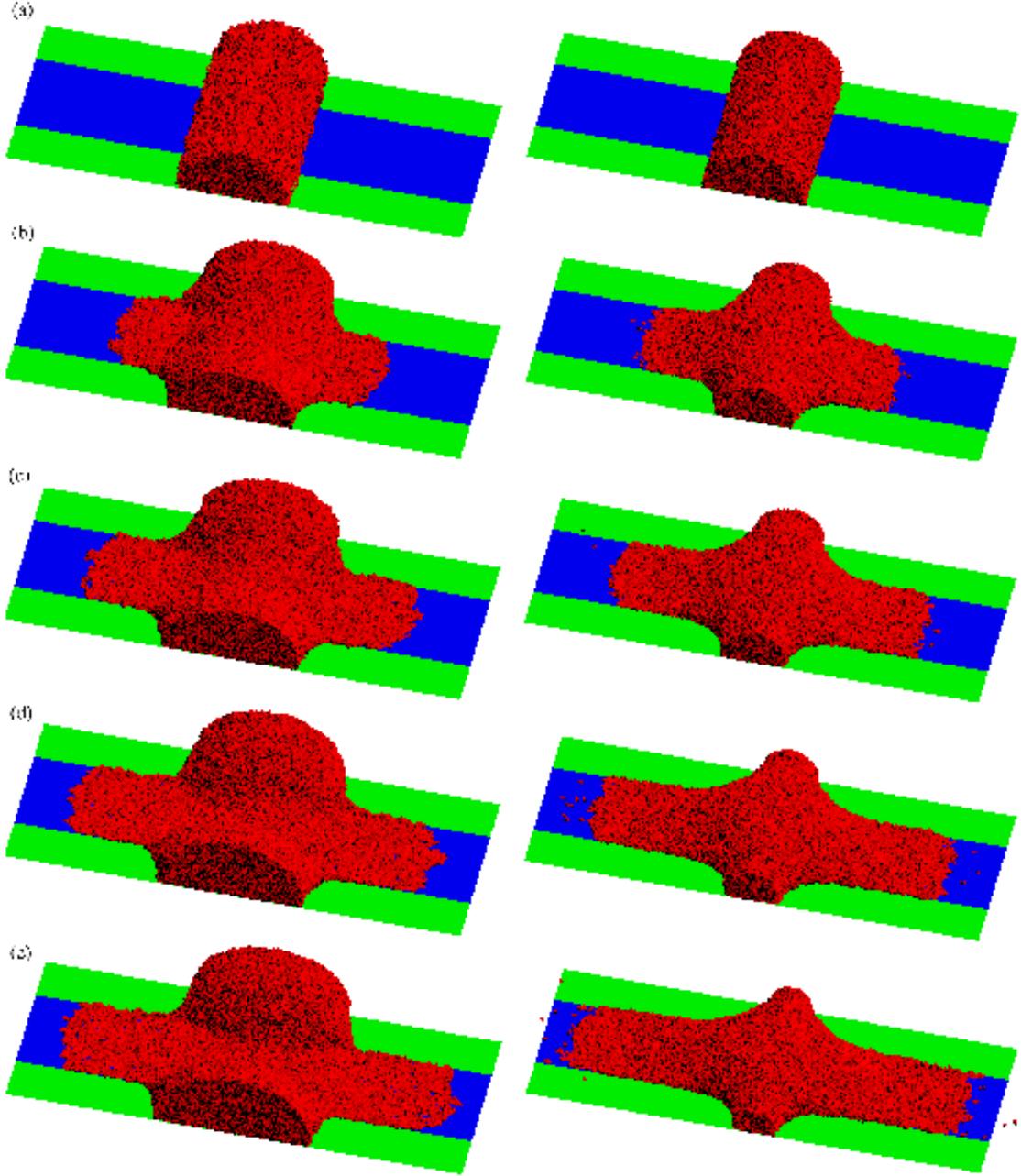}
\end{center}
\caption{\label{cap:drop_100t} (color)
Sequence of simulation snapshots for the $N=100$ (left) and $N=10$ (right)
droplets for $\lambda=200\sigma$
for (a) $t=0$,(b) $t=12,000$, (c) $t=24,000$, (d) $t=36,000$, (e) $t=48,000\tau$.
For $N=100$, in non-wetting regions $\varepsilon_w=1.0\varepsilon$ (green), while in wetting
regions $\varepsilon_w=2.5\varepsilon$ (blue).  For $N=10$, the corresponding
values are $\varepsilon_w=0.9\varepsilon$ (green)
and $\varepsilon_w=2.0\varepsilon$ (blue).
The substrate is $425\sigma$ long and width $L_y=\lambda$.
}
\end{figure}

Figure 2 presents snapshots of $N=10$ droplets at late time
for varying $\lambda$.
The snapshots are shown at different times chosen
so that the size of the precursor foot, when the foot spreads ahead
of the droplet, is approximately the same size in each snapshot.
In the weak wetting case (Fig.~2, left panels), the
main droplet increases in size only slightly before saturating
for $\lambda\leq 100\sigma$ as shown in Fig.~3.
For $\lambda=30\sigma$, the bulk droplet
saturates at a radius $r_b\cong 60\sigma$ in both the wetting and non-wetting regimes.
As $\lambda$ increases, the radius of the droplet in the non-wetting region decreases
while the radius  of the droplet in the wetting regime increases slightly.
For 
$\lambda\ge 150\sigma$, the main droplet spreads.
Similar to the behavior seen in Fig.~1 for $\lambda=200\sigma$,
the $\lambda=150\sigma$
system in Fig.~2d (left) shows that the cross-section of material in the
non-wetting region shrinks as the bulk droplet in the
wetting region continually spreads.
For $\lambda=150\sigma$,
the radius of the droplet in the wetting regime continues to grow at
late time as shown in Fig.~3.
Between $\lambda=100\sigma$ and $150\sigma$ the gain in
surface energy exceeds the cost of increasing the surface area and
material is able to be drawn continually from the non-wetting
region to the wetting region.
Figure 2 (right) presents $N=10$ results for
the strong wetting case.  Compared to the weak wetting case
where spreading of a
precursor foot was nearly absent for $\lambda=30\sigma$,
Fig.~2a (right) clearly shows precursor spreading.
A separate simulation of the strong wetting
system with $\lambda=20\sigma$ showed
a precursor foot spreads even for this very small feature size.
It is also apparent, especially for smaller $\lambda$, that
the stronger interaction in the wetting region causes more spreading
in the non-wetting region.  The $\lambda=150\sigma$ case for strong
wetting is not shown in Fig.~2; however, similar to the
weak wetting case, the minimum
width on which the bulk droplet spreads is between $\lambda=100$ and
$150\sigma$.  This indicates that the bulk wetting transition is related
more to $\lambda$ than to the strength of interaction in the wetting
region $\epsilon_w$.  For reference, $\lambda=\infty$ is shown for
the strong wetting case in Fig.~2d (right).

\begin{figure}
\begin{center}
\includegraphics[%
  clip,
  width=17.8cm,
  keepaspectratio]{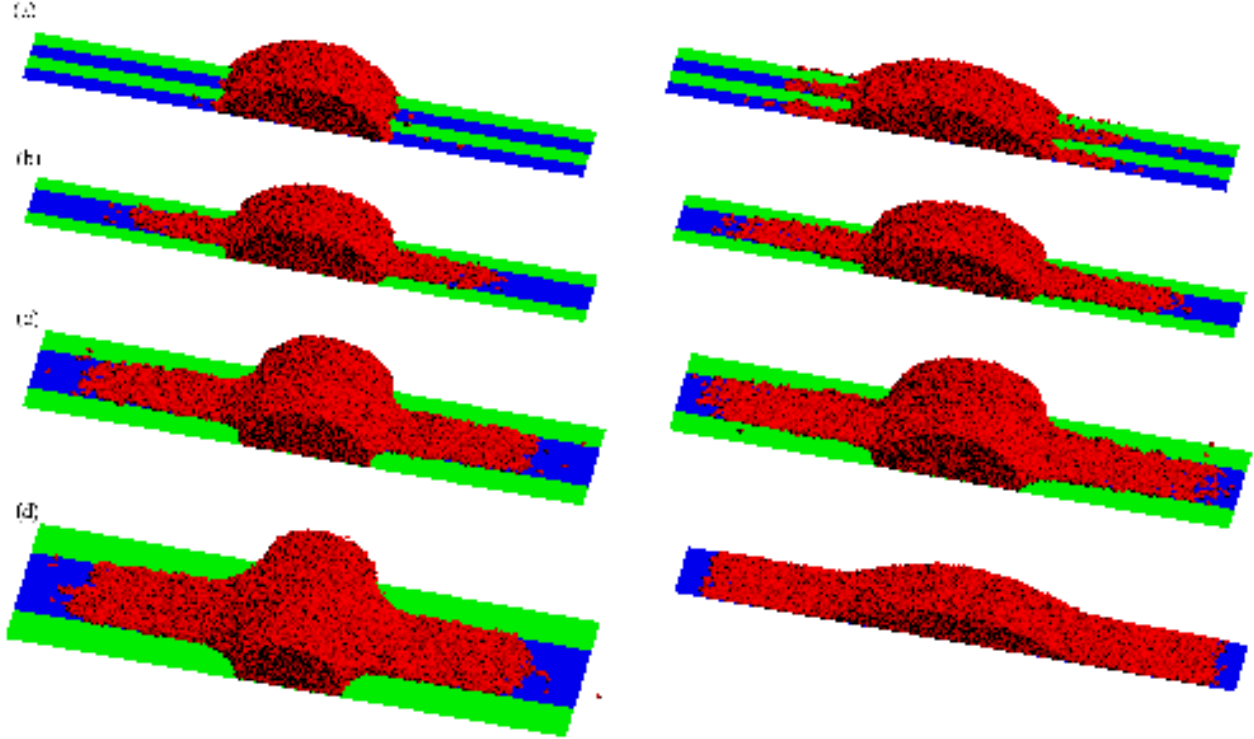}
\end{center}
\caption{\label{cap:drop_10-2} (color)
Simulation snapshots for $N=10$ polymer droplets; in all cases shown, the
interaction strength of the wall in the non-wetting regions (green) is
$\varepsilon_w=0.9\varepsilon$.  For panels on the left, the interaction strength
in the wetting regions (blue) is $\varepsilon_w=2.0\varepsilon$ while
on the right $\varepsilon_w=3.0\varepsilon$.  In panels (a) - (c),
snapshots are presented at $t=80,000\tau$ (left) and $t=30,000\tau$ (right) while
in panel (d) $t=48,000\tau$ (left) and $t=15,000\tau$ (right).
Pattern lengths are (a) $\lambda=30\sigma$,
(b) $\lambda=60\sigma$, (c) $\lambda=100\sigma$, and
(d) $\lambda=150\sigma$ (left) or (d) $\lambda=\infty$ (right).  The
substrate is $450\sigma$ long in each snapshot,
while the width $L_y=\lambda$ except for $\lambda=30\sigma$ and $\lambda=\infty$,
where $L_y=60\sigma$.
}
\end{figure}

\begin{figure}
\begin{center}
\includegraphics[%
  clip,
  width=8cm,
  keepaspectratio]{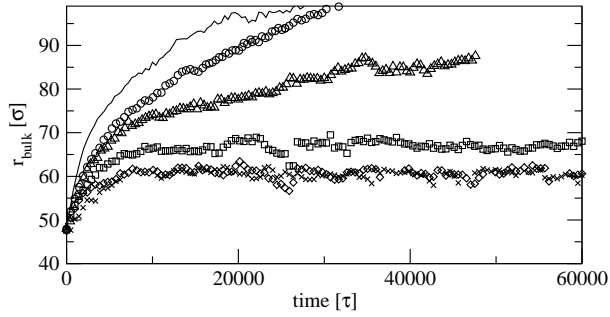}
\end{center}
\caption{\label{cap:bulk_rb} 
Radii of the bulk droplets for $N=10$ in the center of
the wetting regions for
the weak wetting case $\varepsilon_w=0.9$ and $2.0\varepsilon$ for
$\lambda=30\sigma$ (crosses),
$\lambda=60\sigma$  (diamonds),
$\lambda=100\sigma$  (squares),
$\lambda=150\sigma$  (triangles),
$\lambda=200\sigma$  (circles),
and $\lambda=\infty$  (solid).
}
\end{figure}

$N=100$ results for different $\lambda$ are shown in Fig.~4.
At $\lambda=60\sigma$, which is  about 4 times the
average end-to-end  distance of the chain in the melt,
the droplet spreads at early time to a radius $r_b\cong 80\sigma$
before reaching its equilibrium size.  This final size is
comparable to what is observed
for a homogenous surface with a mean surface interaction strength,
$\bar\varepsilon_w\simeq 1.75\varepsilon$.
For $\lambda=100\sigma$ (Fig.~4b) and $200\sigma$ (Fig.~1),  the
bulk radii in the wetting region
saturate at lower values, $r_b\approx 70\sigma$.
These data demonstrate that, for sufficiently small feature size,
a mean interaction description is fairly accurate but becomes
less so for increasing feature size.
For $\lambda\ge 60\sigma$ a precursor foot spreads ahead of the droplet
while for $\lambda\le 30\sigma$ no foot is observed.
The width $\lambda$ above which the main droplet spreads on the
lyophilic strip is larger than
we can access with present computational resources.
For comparison purposes, Fig.~4c presents data for $\lambda=\infty$
and it is obvious that the main droplet spreads
for this limit of pattern size. 

\begin{figure}
\begin{center}
\includegraphics[%
  clip,
  width=8.9cm,
  keepaspectratio]{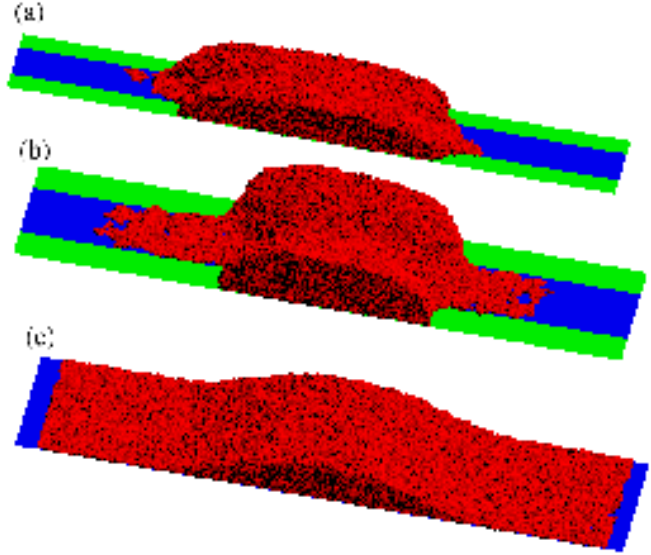}
\end{center}
\caption{\label{cap:drop_100} (color)
Simulation snapshots for $N=100$ polymer droplets
with $\varepsilon_w=1.0\varepsilon$ (green) and
$\varepsilon_w=2.5\varepsilon$ (blue)
for (a) $\lambda=60\sigma$ at $t=96,000\tau$,
(b) $\lambda=100\sigma$ at t$=96,000\tau$,
and (c) $\lambda=\infty$ at $t=32,000\tau$.
The substrate is $425\sigma$ long in each snapshot,
while the width $L_y=\lambda$ except for $\lambda=\infty$, where $L_y=60\sigma$.
}
\end{figure}

The contact radius of the precursor foot increases as
$r_{f}^{2}(t)=2D_{\rm eff}t$ for all cases studied.
The effective diffusion constant $D_{\rm eff}$
normalized by its value for the homogeneous wetting case is shown in Fig.~5.
A result observed in all cases is that $D_{\rm eff}$
increases monotonically with $\lambda$.
Of note is an apparent cross-over in observed behavior with decreasing
$\lambda$ for the $N=10$ strong and weak wetting systems.  Data
imply that, at large pattern
width, foot transport for the stronger wetting system is more
affected by the presence of a lyophobic region than it is for the
weaker wetting system.  We hypothesize that this results
from the very aggressive rate of transport for the foot on the
stronger wetting substrate.  The system must supply material
to the foot from the bulk of the droplet.
The rate of foot advancement is large
enough in the strong wetting case that it becomes limited by transport of
material from the non-wetting region
to the wetting region, and on to the foot.
This is only true for large pattern width $\lambda > 60\sigma$
where the foot is comprised of a more significant amount of polymer
material.  Note that foot transport along the lyophilic strip is
observed for $\lambda < 30\sigma$ in the strong wetting
system whereas it is not for the weaker interaction.
In our previous work \cite{HGW:PRL:05}
studying wetting on homogeneous substrates, we found that $D_{\rm eff}$
increased with the size of the initial droplet as $R_0^{1/2}$. While we have
not varied the size of the initial droplet in this study, we expect a similar dependence
of $D_{\rm eff}$ on the initial size of the droplet.

\begin{figure}
\begin{center}
\includegraphics[%
  clip,
  width=8cm,
  keepaspectratio]{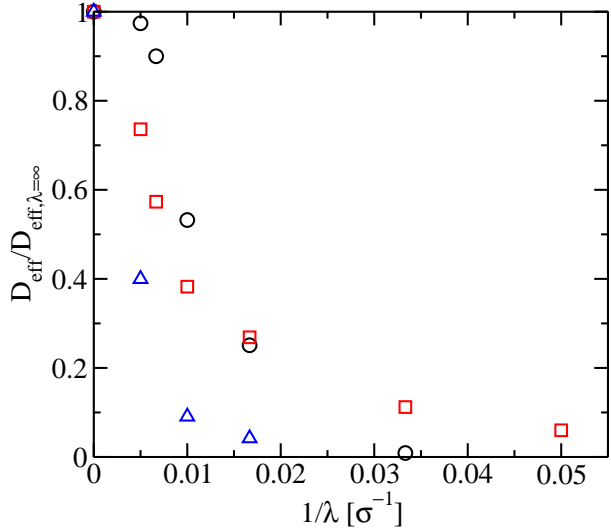}
\end{center}
\caption{\label{cap:ddiff}
Normalized effective diffusion coefficient for the precursor foot as a function of the
surface pattern wavelength $\lambda$ for
$N=10$, $\varepsilon_w=0.9\varepsilon$ and $2.0\varepsilon$ (circles),
$N=10$, $\varepsilon_w=0.9\varepsilon$ and $3.0\varepsilon$ (squares),
$N=100$, $\varepsilon_w=1.0\varepsilon$ and $2.5\varepsilon$ (triangles).
For $R_0=50\sigma$, the values for $D_{\rm eff,\lambda=\infty}=0.17$ (circles),
$0.75$ (squares), and $0.33\sigma^2/\tau$ (triangles).
}
\end{figure}

\section{Conclusion}
\label{sec:conclusions}

Chemically patterning a surface offers a unique opportunity to control the
spreading of liquid nanodroplets. Here we have presented simulation
results which demonstrate the interplay
between the pattern width $\lambda$, the length
of the polymer chain $N$, and the interaction
strength of the lyophilic strips in determining spreading behavior.
We find that there exists a minimum $\lambda$,
which is a function of $N$, below which
spreading of the main droplet does not occur.
For $N=10$, a bulk spreading transition with
increasing $\lambda$ was observed, while
for $N=100$, the transition occurs for $\lambda$ larger than is
presently accessible.
In a number of cases, a precursor foot is observed to
spread even
when the main droplet does not spread.
The observed effective diffusion constant of this
precursor foot decreases monotonically with $\lambda$.
Future work will reveal the dependence of observed kinetics on
the initial size of the droplet and specify the
relation between $N$, $\lambda$, and wetting transition.
 
\begin{acknowledgments}
Sandia is a multiprogram laboratory operated by Sandia Corporation, a Lockheed
Martin Company, for the United States Department of Energy's National
Nuclear Security Administration under contract DE-AC04-94AL85000.
\end{acknowledgments}
  
\newpage


\begin{thebibliography}{25}
\expandafter\ifx\csname natexlab\endcsname\relax\def\natexlab#1{#1}\fi
\expandafter\ifx\csname bibnamefont\endcsname\relax
  \def\bibnamefont#1{#1}\fi
\expandafter\ifx\csname bibfnamefont\endcsname\relax
  \def\bibfnamefont#1{#1}\fi
\expandafter\ifx\csname citenamefont\endcsname\relax
  \def\citenamefont#1{#1}\fi
\expandafter\ifx\csname url\endcsname\relax
  \def\url#1{\texttt{#1}}\fi
\expandafter\ifx\csname urlprefix\endcsname\relax\def\urlprefix{URL }\fi
\providecommand{\bibinfo}[2]{#2}
\providecommand{\eprint}[2][]{\url{#2}}

\bibitem[{\citenamefont{Heslot et~al.}(1989{\natexlab{a}})\citenamefont{Heslot,
  Fraysse, and Cazabat}}]{HFC:Nat:89}
\bibinfo{author}{\bibfnamefont{F.}~\bibnamefont{Heslot}},
  \bibinfo{author}{\bibfnamefont{N.}~\bibnamefont{Fraysse}}, \bibnamefont{and}
  \bibinfo{author}{\bibfnamefont{A.~M.} \bibnamefont{Cazabat}},
  \bibinfo{journal}{Nature} \textbf{\bibinfo{volume}{338}},
  \bibinfo{pages}{640} (\bibinfo{year}{1989}{\natexlab{a}}).

\bibitem[{\citenamefont{Heslot et~al.}(1989{\natexlab{b}})\citenamefont{Heslot,
  Cazabat, and Levinson}}]{HCL:PRL:89}
\bibinfo{author}{\bibfnamefont{F.}~\bibnamefont{Heslot}},
  \bibinfo{author}{\bibfnamefont{A.~M.} \bibnamefont{Cazabat}},
  \bibnamefont{and} \bibinfo{author}{\bibfnamefont{P.}~\bibnamefont{Levinson}},
  \bibinfo{journal}{Phys. Rev. Lett.} \textbf{\bibinfo{volume}{62}},
  \bibinfo{pages}{1286} (\bibinfo{year}{1989}{\natexlab{b}}).

\bibitem[{\citenamefont{Heslot et~al.}(1990)\citenamefont{Heslot, Cazabat,
  Levinson, and Fraysse}}]{HCL:PRL:90}
\bibinfo{author}{\bibfnamefont{F.}~\bibnamefont{Heslot}},
  \bibinfo{author}{\bibfnamefont{A.~M.} \bibnamefont{Cazabat}},
  \bibinfo{author}{\bibfnamefont{P.}~\bibnamefont{Levinson}}, \bibnamefont{and}
  \bibinfo{author}{\bibfnamefont{N.}~\bibnamefont{Fraysse}},
  \bibinfo{journal}{Phys. Rev. Lett.} \textbf{\bibinfo{volume}{65}},
  \bibinfo{pages}{599} (\bibinfo{year}{1990}).

\bibitem[{\citenamefont{Xu et~al.}(2004)\citenamefont{Xu, Shirvanyants, Beers,
  Matyjaszewski, Rubinstein, and Sheiko}}]{XSB:PRL:04}
\bibinfo{author}{\bibfnamefont{H.}~\bibnamefont{Xu}},
  \bibinfo{author}{\bibfnamefont{D.}~\bibnamefont{Shirvanyants}},
  \bibinfo{author}{\bibfnamefont{K.}~\bibnamefont{Beers}},
  \bibinfo{author}{\bibfnamefont{K.}~\bibnamefont{Matyjaszewski}},
  \bibinfo{author}{\bibfnamefont{M.}~\bibnamefont{Rubinstein}},
  \bibnamefont{and} \bibinfo{author}{\bibfnamefont{S.~S.}
  \bibnamefont{Sheiko}}, \bibinfo{journal}{Phys. Rev. Lett.}
  \textbf{\bibinfo{volume}{93}}, \bibinfo{pages}{206103}
  (\bibinfo{year}{2004}).

\bibitem[{\citenamefont{Nieminen et~al.}(1992)\citenamefont{Nieminen, Abraham,
  Karttunen, and Kaski}}]{NAK:PRL:92}
\bibinfo{author}{\bibfnamefont{J.~A.} \bibnamefont{Nieminen}},
  \bibinfo{author}{\bibfnamefont{D.~B.} \bibnamefont{Abraham}},
  \bibinfo{author}{\bibfnamefont{M.}~\bibnamefont{Karttunen}},
  \bibnamefont{and} \bibinfo{author}{\bibfnamefont{K.}~\bibnamefont{Kaski}},
  \bibinfo{journal}{Phys. Rev. Lett.} \textbf{\bibinfo{volume}{69}},
  \bibinfo{pages}{124} (\bibinfo{year}{1992}).

\bibitem[{\citenamefont{D'Ortona et~al.}(1996)\citenamefont{D'Ortona,
  De~Coninck, Koplik, and Banavar}}]{OCK:PRE:96}
\bibinfo{author}{\bibfnamefont{U.}~\bibnamefont{D'Ortona}},
  \bibinfo{author}{\bibfnamefont{J.}~\bibnamefont{De~Coninck}},
  \bibinfo{author}{\bibfnamefont{J.}~\bibnamefont{Koplik}}, \bibnamefont{and}
  \bibinfo{author}{\bibfnamefont{J.~R.} \bibnamefont{Banavar}},
  \bibinfo{journal}{Phys. Rev. E} \textbf{\bibinfo{volume}{53}},
  \bibinfo{pages}{562} (\bibinfo{year}{1996}).

\bibitem[{\citenamefont{Bekink et~al.}(1996)\citenamefont{Bekink, Karaborni,
  Verbist, and Esselink}}]{BKV:PRL:96}
\bibinfo{author}{\bibfnamefont{S.}~\bibnamefont{Bekink}},
  \bibinfo{author}{\bibfnamefont{S.}~\bibnamefont{Karaborni}},
  \bibinfo{author}{\bibfnamefont{G.}~\bibnamefont{Verbist}}, \bibnamefont{and}
  \bibinfo{author}{\bibfnamefont{K.}~\bibnamefont{Esselink}},
  \bibinfo{journal}{Phys. Rev. Lett.} \textbf{\bibinfo{volume}{76}},
  \bibinfo{pages}{3766} (\bibinfo{year}{1996}).

\bibitem[{\citenamefont{Heine et~al.}(2003)\citenamefont{Heine, Grest, and
  Webb~III}}]{HGW:PRE:03}
\bibinfo{author}{\bibfnamefont{D.~R.} \bibnamefont{Heine}},
  \bibinfo{author}{\bibfnamefont{G.~S.} \bibnamefont{Grest}}, \bibnamefont{and}
  \bibinfo{author}{\bibfnamefont{E.~B.} \bibnamefont{Webb~III}},
  \bibinfo{journal}{Phys. Rev. E} \textbf{\bibinfo{volume}{68}},
  \bibinfo{pages}{061603} (\bibinfo{year}{2003}).

\bibitem[{\citenamefont{Heine et~al.}(2004)\citenamefont{Heine, Grest, and
  Webb~III}}]{HGW:PRE:04}
\bibinfo{author}{\bibfnamefont{D.~R.} \bibnamefont{Heine}},
  \bibinfo{author}{\bibfnamefont{G.~S.} \bibnamefont{Grest}}, \bibnamefont{and}
  \bibinfo{author}{\bibfnamefont{E.~B.} \bibnamefont{Webb~III}},
  \bibinfo{journal}{Phys. Rev. E} \textbf{\bibinfo{volume}{70}},
  \bibinfo{pages}{011606} (\bibinfo{year}{2004}).

\bibitem[{\citenamefont{Heine et~al.}(2005)\citenamefont{Heine, Grest, and
  Webb~III}}]{HGW:PRL:05}
\bibinfo{author}{\bibfnamefont{D.~R.} \bibnamefont{Heine}},
  \bibinfo{author}{\bibfnamefont{G.~S.} \bibnamefont{Grest}}, \bibnamefont{and}
  \bibinfo{author}{\bibfnamefont{E.~B.} \bibnamefont{Webb~III}},
  \bibinfo{journal}{Phys. Rev. Lett.} \textbf{\bibinfo{volume}{95}},
  \bibinfo{pages}{107801} (\bibinfo{year}{2005}).

\bibitem[{\citenamefont{Xia and Whitesides}(1998)}]{XW:ARMS:98}
\bibinfo{author}{\bibfnamefont{Y.}~\bibnamefont{Xia}} \bibnamefont{and}
  \bibinfo{author}{\bibfnamefont{G.~M.} \bibnamefont{Whitesides}},
  \bibinfo{journal}{Annu. Rev. Mater. Res.} \textbf{\bibinfo{volume}{28}},
  \bibinfo{pages}{153} (\bibinfo{year}{1998}).

\bibitem[{\citenamefont{Lipowsky}(2001)}]{Lip:COCIS:01}
\bibinfo{author}{\bibfnamefont{R.}~\bibnamefont{Lipowsky}},
  \bibinfo{journal}{Curr. Opin. Colloid Interface Sci.}
  \textbf{\bibinfo{volume}{6}}, \bibinfo{pages}{40} (\bibinfo{year}{2001}).

\bibitem[{\citenamefont{Darhuber and Troian}(2005)}]{DT:ARFM:05}
\bibinfo{author}{\bibfnamefont{A.~A.} \bibnamefont{Darhuber}} \bibnamefont{and}
  \bibinfo{author}{\bibfnamefont{S.~M.} \bibnamefont{Troian}},
  \bibinfo{journal}{Ann. Rev. Fluid Mech.} \textbf{\bibinfo{volume}{37}},
  \bibinfo{pages}{425} (\bibinfo{year}{2005}).

\bibitem[{\citenamefont{Gau et~al.}(1999)\citenamefont{Gau, Herminghaus, Lenz,
  and Lipowsky}}]{GHLL:SCI:99}
\bibinfo{author}{\bibfnamefont{H.}~\bibnamefont{Gau}},
  \bibinfo{author}{\bibfnamefont{S.}~\bibnamefont{Herminghaus}},
  \bibinfo{author}{\bibfnamefont{P.}~\bibnamefont{Lenz}}, \bibnamefont{and}
  \bibinfo{author}{\bibfnamefont{R.}~\bibnamefont{Lipowsky}},
  \bibinfo{journal}{Science} \textbf{\bibinfo{volume}{283}},
  \bibinfo{pages}{46} (\bibinfo{year}{1999}).

\bibitem[{\citenamefont{Darhuber et~al.}(2000)\citenamefont{Darhuber, Troian,
  Miller, and Wagner}}]{DTMW:JAP:00}
\bibinfo{author}{\bibfnamefont{A.~A.} \bibnamefont{Darhuber}},
  \bibinfo{author}{\bibfnamefont{S.~M.} \bibnamefont{Troian}},
  \bibinfo{author}{\bibfnamefont{S.~M.} \bibnamefont{Miller}},
  \bibnamefont{and} \bibinfo{author}{\bibfnamefont{S.}~\bibnamefont{Wagner}},
  \bibinfo{journal}{J. Appl. Phys.} \textbf{\bibinfo{volume}{87}},
  \bibinfo{pages}{7768} (\bibinfo{year}{2000}).

\bibitem[{\citenamefont{Brinkmann and Lipowsky}(2002)}]{BL:JAP:02}
\bibinfo{author}{\bibfnamefont{M.}~\bibnamefont{Brinkmann}} \bibnamefont{and}
  \bibinfo{author}{\bibfnamefont{R.}~\bibnamefont{Lipowsky}},
  \bibinfo{journal}{J. Appl. Phys.} \textbf{\bibinfo{volume}{92}},
  \bibinfo{pages}{4296} (\bibinfo{year}{2002}).

\bibitem[{\citenamefont{L{\`e}opold{\`e}s
  et~al.}(2003)\citenamefont{L{\`e}opold{\`e}s, Dupuis, Bucknall, and
  Yeomans}}]{LDBY:LAN:03}
\bibinfo{author}{\bibfnamefont{J.}~\bibnamefont{L{\`e}opold{\`e}s}},
  \bibinfo{author}{\bibfnamefont{A.}~\bibnamefont{Dupuis}},
  \bibinfo{author}{\bibfnamefont{D.~G.} \bibnamefont{Bucknall}},
  \bibnamefont{and} \bibinfo{author}{\bibfnamefont{J.~M.}
  \bibnamefont{Yeomans}}, \bibinfo{journal}{Langmuir}
  \textbf{\bibinfo{volume}{19}}, \bibinfo{pages}{9818} (\bibinfo{year}{2003}).

\bibitem[{\citenamefont{Dupuis and Yeomans}(2005)}]{DY:LAN:05}
\bibinfo{author}{\bibfnamefont{A.}~\bibnamefont{Dupuis}} \bibnamefont{and}
  \bibinfo{author}{\bibfnamefont{J.~M.} \bibnamefont{Yeomans}},
  \bibinfo{journal}{Langmuir} \textbf{\bibinfo{volume}{21}},
  \bibinfo{pages}{2624} (\bibinfo{year}{2005}).

\bibitem[{\citenamefont{Yaneva et~al.}(2005)\citenamefont{Yaneva, Milchev, and
  Binder}}]{YMB:JCP:05}
\bibinfo{author}{\bibfnamefont{J.}~\bibnamefont{Yaneva}},
  \bibinfo{author}{\bibfnamefont{A.}~\bibnamefont{Milchev}}, \bibnamefont{and}
  \bibinfo{author}{\bibfnamefont{K.}~\bibnamefont{Binder}},
  \bibinfo{journal}{J. Chem. Phys.} \textbf{\bibinfo{volume}{123}},
  \bibinfo{pages}{0} (\bibinfo{year}{2005}).

\bibitem[{\citenamefont{Dietrich et~al.}(2005)\citenamefont{Dietrich, Popescu,
  and Rauscher}}]{DPR:JOP:05}
\bibinfo{author}{\bibfnamefont{S.}~\bibnamefont{Dietrich}},
  \bibinfo{author}{\bibfnamefont{M.~N.} \bibnamefont{Popescu}},
  \bibnamefont{and} \bibinfo{author}{\bibfnamefont{M.}~\bibnamefont{Rauscher}},
  \bibinfo{journal}{J. Phys.: Condens. Mater.} \textbf{\bibinfo{volume}{17}},
  \bibinfo{pages}{S577} (\bibinfo{year}{2005}).

\bibitem[{\citenamefont{Dupuis et~al.}(2005)\citenamefont{Dupuis,
  L{\`e}opold{\`e}s, Bucknall, and Yeomans}}]{DLBY:CM:05}
\bibinfo{author}{\bibfnamefont{A.}~\bibnamefont{Dupuis}},
  \bibinfo{author}{\bibfnamefont{J.}~\bibnamefont{L{\`e}opold{\`e}s}},
  \bibinfo{author}{\bibfnamefont{D.~G.} \bibnamefont{Bucknall}},
  \bibnamefont{and} \bibinfo{author}{\bibfnamefont{J.~M.}
  \bibnamefont{Yeomans}}, \bibinfo{journal}{Cond-mat/} p.
  \bibinfo{pages}{0507335} (\bibinfo{year}{2005}).

\bibitem[{\citenamefont{P{\"u}tz et~al.}(2000)\citenamefont{P{\"u}tz, Kremer,
  and Grest}}]{PKG:EPL:00}
\bibinfo{author}{\bibfnamefont{M.}~\bibnamefont{P{\"u}tz}},
  \bibinfo{author}{\bibfnamefont{K.}~\bibnamefont{Kremer}}, \bibnamefont{and}
  \bibinfo{author}{\bibfnamefont{G.~S.} \bibnamefont{Grest}},
  \bibinfo{journal}{Europhys. Lett.} \textbf{\bibinfo{volume}{49}},
  \bibinfo{pages}{735} (\bibinfo{year}{2000}).

\bibitem[{\citenamefont{Kremer and Grest}(1990)}]{KG:JCP:90}
\bibinfo{author}{\bibfnamefont{K.}~\bibnamefont{Kremer}} \bibnamefont{and}
  \bibinfo{author}{\bibfnamefont{G.~S.} \bibnamefont{Grest}},
  \bibinfo{journal}{J. Chem. Phys.} \textbf{\bibinfo{volume}{92}},
  \bibinfo{pages}{5057} (\bibinfo{year}{1990}).

\bibitem[{\citenamefont{Braun and Peyrard}(2001)}]{BP:PRE:01}
\bibinfo{author}{\bibfnamefont{O.~M.} \bibnamefont{Braun}} \bibnamefont{and}
  \bibinfo{author}{\bibfnamefont{M.}~\bibnamefont{Peyrard}},
  \bibinfo{journal}{Phys. Rev. E} \textbf{\bibinfo{volume}{63}},
  \bibinfo{pages}{046110} (\bibinfo{year}{2001}).

\bibitem[{\citenamefont{Plimpton}(1995)}]{P:JCP:95}
\bibinfo{author}{\bibfnamefont{S.}~\bibnamefont{Plimpton}},
  \bibinfo{journal}{J. Comput. Phys.} \textbf{\bibinfo{volume}{117}},
  \bibinfo{pages}{1} (\bibinfo{year}{1995}).

\end{thebibliography}

\end{document}